**Title**: The Gaussian-Linear Hidden Markov model: a Python package

Diego Vidaurre[a, b], Laura Masaracchia[a], Nick Y. Larsen[a], Lenno R.P.T Ruijters[a], Sonsoles Alonso[a], Christine Ahrends[a], and Mark W. Woolrich[b]

a - Center of Functionally Integrative Neuroscience, Dept. of Clinical Medicine, Aarhus University, Denmark
b - Oxford Centre for Human Brain Analysis, Psychiatry Department, Oxford University, Oxford, UK

**Abstract**: We propose the Gaussian-Linear Hidden Markov model (GLHMM), a generalisation of different types of HMMs commonly used in neuroscience. In short, the GLHMM is a general framework where linear regression is used to flexibly parameterise the Gaussian state distribution, thereby accommodating a wide range of uses —including unsupervised, encoding and decoding models. GLHMM is implemented as a Python toolbox with an emphasis on statistical testing and out-of-sample prediction —i.e. aimed at finding and characterising brain-behaviour associations. The toolbox uses a stochastic variational inference approach, enabling it to handle large data sets at reasonable computational time. The approach can be applied to several data modalities, including animal recordings or non-brain data, and applied over a broad range of experimental paradigms. For demonstration, we show examples with fMRI, electrocorticography, magnetoencephalography and pupillometry.

## Introduction

Brain activity is highly multidimensional and complex. Finding structure within this complexity is crucial to establish meaningful associations to behaviour. Consequently, in recent years the Hidden Markov Model (HMM), with its capacity for finding dynamic network configurations in a time-resolved manner, has emerged as a general family of models that can be applied to a broad array of scientific questions and data modalities (Vidaurre et al., 2017; Stevner et al., 2019; Vidaurre et al., 2018a; Higgins et al., 2021). For this, the HMM typically represents a signal with rich properties using a reduced number of components (i.e. the HMM states), such that each component explains a specific subset of the signal. As opposed to principal or independent component analysis, these components can capture advanced aspects of the signal such as specific patterns of correlation or spectral properties (Vidaurre et al., 2016). Importantly, even when assumptions made by the HMM may be questionable in brain data, (such as the assumption of discrete brain states), it has proven to be a valuable and parsimonious description of dynamic processes in the data (Stevner et al., 2019; Higgins et al., 2021).

More specifically, the HMM is a hierarchical probabilistic model that uses a set of states that switch on and off (typically over time) as a richer description of the data beyond simple (temporal) averaging. In the context of the HMM, a state describes the data (at the time points when that state is "active") using state-specific probability distributions that can be customised to suit the characteristics of different data modalities. These state distributions are sometimes also referred to as the emission distributions or observation models. For example, a commonly used approach is to use Gaussian state distributions that are multivariate over channels or brain regions. This approach can be adapted to describe specific patterns of signal amplitude and/or frequency, functional connectivity, etc. A non-exhaustive list of state distributions, all based on the Gaussian distribution, that have been previously used in neuroscience applications are:

- The Gaussian distribution, on fMRI and other neuroimaging modalities, where the states capture spatial information, i.e. the average activation pattern and functional connectivity (covariance) of the BOLD signal between areas across the whole brain (Vidaurre et al., 2017; Stevner et al., 2019; Vidaurre et al., 2018a).
- A Wishart distribution on fMRI, if we wish to focus on changes in functional connectivity (i.e. covariance; Vidaurre et al., 2021; Ahrends et al., 2022; Alonso & Vidaurre, 2023).
- The time-delay embedded distribution on electrophysiological data (e.g. MEG or EEG), which, based on the Gaussian distribution, can capture spectral modulations on high-dimensional data (Vidaurre et al., 2018b).
- The autoregressive model, also used on electrophysiological data, which, with a larger number of parameters than the time-delay embedded model, offers a more detailed description of the spectral aspects in the data. This is therefore more appropriate for temporally richer modalities like local field potentials (LFP; Vidaurre et al., 2016; Masaracchia et al., 2023).
- A decoding model to describe the changing relation between brain activity and ongoing stimuli by explicitly including task information in the model. Specifically, this is a regression model where the brain data act as independent variables, and the stimuli as dependent variables (Vidaurre et al., 2019).
- An encoding model, which also describes the relation between brain activity and ongoing stimuli, but where the focus is on spatial interpretation. Here, the brain data are the dependent variables, and the stimuli are the independent variables (Higgins et al., 2022).

Each one of these distributions yields a different variety of the HMM. In this paper, we propose the Gaussian-Linear Hidden Markov Model (GLHMM), a generalisation of all the above. Also, we present a Python toolbox available on PyPI[1] with a focus on routines to relate the models to experimental conditions, observed behaviour, and subject traits via statistical testing and out-of-sample prediction. Although here we focus on functional neuroimaging, including electrophysiological data, the model can be applied to many other data types, both neural and non-neural (e.g. purely behavioural data). For very large datasets, regardless of the data modality, the inference of the model parameters can be configured to make use of stochastic learning for computational efficiency and less demanding memory usage. Accompanying this paper, extensive software documentation and example Jupyter notebooks are available[2].

---

[1] https://test.pypi.org/project/glhmm/; https://github.com/vidaurre/glhmm
[2] https://github.com/vidaurre/glhmm/tree/main/docs

# An application example: the synergy between neural oscillations and neural firing

The following example illustrates the motivation behind the Gaussian-Linear Hidden Markov Model, which we will explain in detail in the following sections. Two outstanding questions in systems neuroscience are: 1) how the shape and amplitude of neural oscillations relate to the spiking of individual neurons, and 2) how the complex structure of this neuronal activity relates to behaviour. These two questions have most often been approached separately, but a few exceptions exist that have greatly advanced our understanding of brain function. One is the well-known phenomenon of phase precession, where specific phases of theta induce a preference of place hippocampal pyramidal neurons to fire, depending on the spatial location of the animal with respect to the neurons' place field (Moser et al., 2008).

Here, we show how the GLHMM framework can be used to investigate, from a data-driven perspective, how the relationship between multi-neuron spiking activity and local field potential (LFP) oscillations varies with behaviour. Specifically, we consider an existing data set with electrophysiological recordings obtained from the pyramidal cell layer of the dorsal CA1 hippocampus region in rats, where the task involved recognition of sequences of odours (Shahbaba et al., 2022). We ran two types of the HMM: one where states were purely defined in terms of patterns of average firing (i.e., a standard HMM, where states are Gaussian distributions on the spike densities); and another where states were defined as the relation between the LFP's power in the theta band and the neurons' firing (i.e. a regression-based HMM, where the regressor is the LFP's first principal component and the response is the neurons' spike density; see below for mathematical details). We then used the HMM state time courses (i.e. the probabilities of activation of each state) to predict, across trials and in a cross-validated fashion, a variable related to higher-order cognition: whether the animals were able to recognise if the presented odours were in- or out-of-sequence.

**Figure 1** shows results for two example animals, one per row. **Figure 1A** shows the accuracy of each model, as a function of time. Here, the standard HMM (blue curves) represents a traditional analysis, where we show the relation of the spike densities to behaviour (perceptual success); whereas the regression-based HMM (red curves) represent how the modulation of neuronal spiking by the LFP oscillatory component (i.e. the state-specific regression coefficients) relates to behaviour. While the standard HMM (based purely on the neurons' activity) is often more predictive, we also see that the red curve is at times higher, particularly in the first half of the trials (before the action at 0.0). **Figure 1B** shows the difference between (red minus blue) alongside the statistical significance of the differences (cluster-based permutation testing, significance level = 0.01).

However, do the two models contain complementary information, or is the accuracy of one model nested in the other? **Figure 1C** shows that there are a substantial number of trials where only one of the two methods is able to predict correctly; again, the regression-based HMM was able to decode trials that the standard HMM could not primarily before the action (0.0s). Finally, **Figure 1D** confirms this point by showing the same information before cross-trial-averaging, where red and blue mark, respectively, the trials where only the regression-based, or standard, HMM predicted correctly.

Overall, this analysis shows how we can use the GLHMM framework and its multiple parametrisations to explore different aspects of brain-behaviour associations, providing insights into their mechanistic underpinnings. In the present example, our preliminary analysis hinted at an unexplored three-way relation between mesoscale oscillatory activity, neural spiking and behaviour.

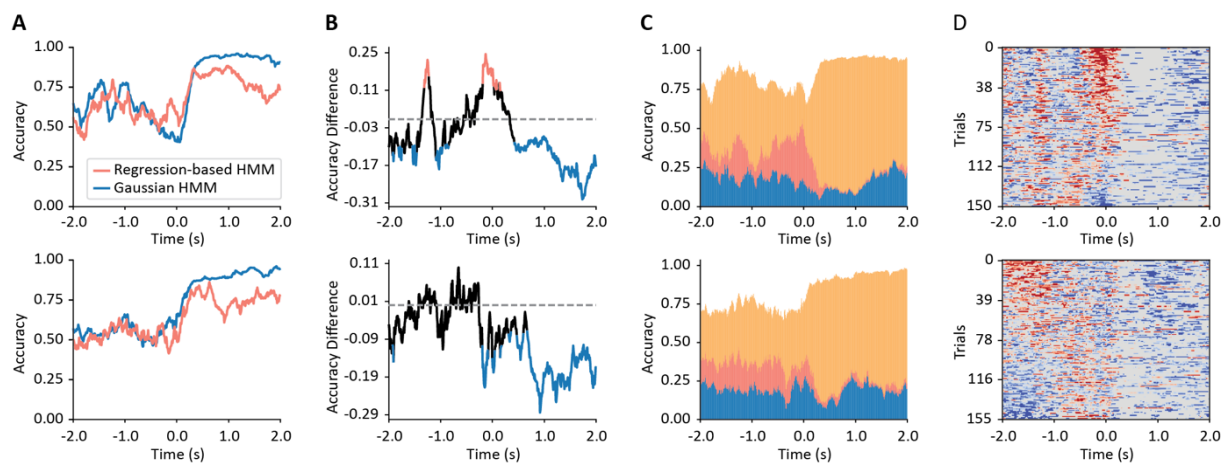

**Figure 1.** Using the GLHMM framework on hippocampal LFP and neuronal spike density data across several neurons. Two types of HMM were run: one considering neural spike densities alone (Gaussian HMM), and the other considering the relation between the power of the LFP oscillations and the spike densities (regression-based HMM). **A)** Accuracies of the Gaussian HMM (blue) and the regression-based HMM (red) in predicting the animals' success in recognising a sequence of odours. **B)** The difference in accuracy between the two approaches (colour represents statistical significance, according to cluster-based corrected permutation testing). **C)** Number of trials where only the Gaussian HMM predicted correctly (blue); where only the regression-based HMM predicted correctly (red); and where both predicted correctly (yellow). **D)** Unaveraged display of which trials were predicted correctly for only one of the models (blue and red respectively); trials were ordered in the Y-axis according to the first principal component of the displayed matrix.

## The Gaussian Linear Hidden Markov model

The HMM is a family of probability distributions commonly used to model time series, or, more generally, sequential data using a discrete number of states. It is also a latent variable model, where the latent variables, $s_t$, express which state is active per time point $t$. The main assumptions of the HMM are two-fold: (i) the data can be reasonably described using a finite number of discrete states $k = 1, \ldots, K$; and (ii) the Markovian assumption, namely conditional independence between $s_{>t}$ and $s_{<t}$ (respectively denoting all latent variables after time point $t$ and all latent variables before time point $t$) given $s_t$. The second assumption implies, mathematically, that the model is sensitive to time dependencies across more than one time point (with the sensitivity decaying exponentially in time). We note that the HMM can still be used to great effect when these assumptions are biophysically questionable, since it still provides a clearly defined and parsimonious description of the dynamic processes in the data.

Based on the Gaussian distribution, the GLHMM is a generalisation of several HMM variations, which we can flexibly tune to different types of data and scientific questions. In its most general form, it models the relation between two sets of time series $X$ and $Y$, both of which can be either univariate or multivariate. Specifically, this is a regression model, where $X$ is the independent variable and $Y$ is the dependent variable. Either $X$ and $Y$, or both, can be measures of brain activity; for example, $X$ could be measurements on the primary visual cortex and $Y$ could be measurements on the associative visual cortex. Either, or both, can be physiological, behavioural or stimulus-related variables. Additionally, these do not need to be single, continuous time series, but can be split into an arbitrary number of segments; for example, they can split into scanning sessions or experimental trials. However, it is required that $X$ and $Y$ have the same length and be simultaneously sampled. For example, the model cannot handle missing values in $X$ or $Y$, if these missing values do not occur in both time series at the same time. The model, illustrated in **Figure 2**, can be described mathematically as follows:

$$Y_t \sim N(\mu^k + X_t \beta^k, \Sigma^k),$$
$$P(s_t = k | s_{t-1} = l) = \theta_{l,k}$$
$$P(s_0 = k) = \pi_k$$

where $Y_t$ follows a (univariate or multivariate) Gaussian distribution, $\mu^k$ is a vector of baseline activity of $y$ when state $k$ is active, $\beta^k$ is a vector of regression coefficients that relates $X$ to $Y$ when state $k$ is active; $\Sigma^k$ is a covariance matrix specific to state $k$; $\theta_{l,k}$ is the probability of transitioning from state $k$ to state $l$; and $\pi_k$ is the probability that each segment of the time series starts with state $k$. The difference with the standard Gaussian HMM is the presence of the $X_t \beta^k$ term.

We set up prior distributions as follows: Gaussian for $\beta^k$ and $\mu^k$, Wishart for $\Sigma^k$, and Dirichlet for $\pi$ and each row of $\theta$. The hyper-parameters of these priors are estimated in the inference.

We define the posterior probability estimates as

$$\gamma_{t,k} := P(s_t = k | s_{>t}, s_{<t}, X, Y)$$
$$\xi_{t,k,l} := P(s_t = k, s_{t-1} = l | s_{>t}, s_{<t-1}, X, Y)$$

where $\gamma_{t,k}$, for $t = 1...T$, $k = 1...K$, are referred to as the state time courses, i.e. the estimated probabilities of state $k$ to be active at time point $t$; and $\xi_{t,k,l}$ are the joint state probabilities for two contiguous time points starting at $t$.

Of note, while the inference yields $\gamma_k$ as probabilities (i.e., summing to 1.0 across states at each time point), a simple modification of the inference can be used to produce the Viterbi path. This has categorical assignments instead of probabilities and can be defined as the most likely sequence of states for every continuous stream of data (e.g., a scanning session). As we will describe later, the Viterbi path is useful for some types of statistical testing. The estimation of all these parameters is done through variational inference (Jordan & Saul, 1999), or, if the data set is too large and the inference becomes too costly, through a variant referred to as stochastic variational inference (Hoffman et al., 2013); see below for an empirical comparison between the two.

The key versatility of our implementation is that $\mu^k$, $\beta^k$ and $\Sigma^k$ can be state-dependent, state-independent (i.e. global - the same for all states), or pre-specified (e.g. $\mu^k$ or $\beta^k$ could be set to zero). For example, if $\beta^k$ is set to zero (i.e. if $X$ is not part of the model because we are analysing only one set of time series), then the state model amounts to a Gaussian distribution; if $\mu^k$ is set to zero then the model is a piecewise regression with $X$ as the independent variable and no intercept; finally, if both $\mu^k$ and $\beta^k$ are modelled, then $Y$ is treated as though it has a component that depends on $X$ and a component that does not. On the other hand, if either $\mu^k$ or $\beta^k$ are state-independent, this means that they are modelled globally, having $\mu^k = \mu$ or $\beta^k = \beta$ for every state $k$ (i.e. for the entire time series). If $\Sigma^k$ is global, then the model assumes that the distribution of the error is the same for the entire time series; whereas if it is state-dependent, we will allow it to change throughout the time series. Furthermore, the covariance matrix $\Sigma^k$ can be either full-matrix, diagonal or pre-specified to the identity matrix. If it is diagonal, the inference will only estimate one variance parameter per column of $Y$; whereas if it is full, it will yield a more complex model, where the covariance between the errors is taken into account in the estimation.

Overall, this general formulation contains three ($\mu^k$ being state-dependent, state-independent/global, or pre-specified) by three ($\beta^k$ being state-dependent, state-independent/global, or pre-specified) by three ($\Sigma^k$ being state-dependent, state-independent/global, or pre-specified to the identity) by two ($\Sigma^k$ being full or diagonal, whenever it is not pre-specified to the identity) possibilities. As illustrated in **Figure 2**, this amounts to 33 different models, after excluding 12 trivial cases where no variable is state-dependent. The appropriate configuration to use in any given situation will depend on the data modality at hand, the scientific question, and which aspects of the data we wish to capture.

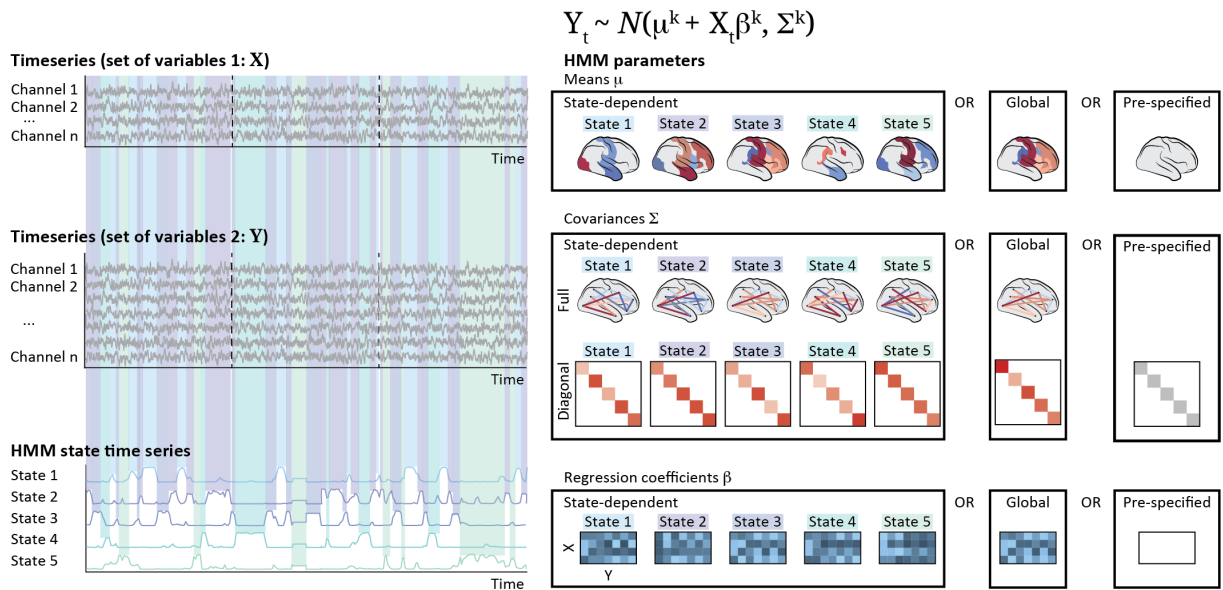

**Figure 2.** Schematic of the GLHMM, with an illustration of the modelling choices. The variables contained in X and Y are generically referred to as channels, but they can represent any type of data.

**Dual estimation**

Analogous to the approach of dual regression, which is used following group independent component analysis to get subject-level ICA representations (Beckmann et al., 2009), a process called dual estimation is used here to obtain subject-level estimations of the HMM parameters (Vidaurre et al., 2021). That is, whereas the state time courses were already subject-specific in the group-level estimation, dual estimation also produces state distribution parameters (i.e. $\mu^k$, $\beta^k$ and $\Sigma^k$) as well as transition and initial probabilities for each subject.

The same idea can be used to obtain session- or even trial-level estimates. This is achieved by rerunning the inference for each subject separately using the state time courses from the group-level estimation, which are not allowed to update in this second step. While an alternative route would be to fit a separate HMM on each subject, the dual estimation approach allows us to straightforwardly have comparable estimations across subjects, for instance such that state 1 always refers to the same network even if it slightly varies across subjects. This is important for the purposes of statistical testing and out-of-sample prediction, as described in later sections.

**Model complexity: number of states and temporal regularisation**

After establishing which one of the abovementioned 33 model variants is going to be used, which primarily will depend on the question and data modality at hand, we need to decide on the level of complexity of the model. In the literature of mixture and latent models (such as the HMM), this means choosing the number of latent components or states. Here, we introduce temporal regularisation as a second way to control the complexity of the model.

Temporal regularisation is performed by adjusting the prior probability of staying in the current state (a Dirichlet prior), making it harder or easier to switch states. Ultimately, large increases in this parameter lead to state pruning, when one or more states stop being visited. This makes this parameter complementary to the number of states, as another way to control the complexity of the model.

Regarding the selection of the number of states, the focus is often to determine the number of states supported by the available data (Vidaurre, 2023). This is consistent with the idea that the HMM is not typically expected to reflect biophysical reality, but instead to provide a useful description of the data. For this, it is sometimes useful to have a quantitative metric to rank different levels of complexity. This is usually achieved by comparing the free energy between HMMs with different numbers of states. The free energy approximates the Bayesian model evidence that balances accuracy (in terms of the likelihood function) with simplicity (in terms of the divergence between the estimated posteriors and the priors). However, in many neuroscience applications we do not even want to choose the number of HMM states based on what is optimal from a Bayesian statistical perspective; instead, the choice is often driven by what makes for a desirable level of detail to provide a useful description. For example, even if the data support more states, it might be convenient to use fewer for a more parsimonious, interpretative description. Another useful approach for model selection is half-split reproducibility, where we fit the model independently on two randomly chosen halves of the data and compare the state distributions across the two runs (Vidaurre et al., 2017).

A useful metric of the resulting complexity of the estimated model is the entropy of the states' fractional occupancies (FO), which are defined as the proportion of time spent in each state $k$. Mathematically, the FO entropy $H$ is defined as

$$H = -\sum_{k} FO_k \log(FO_k)$$

**Model comparison example**

We next show an example of model comparison where we aim to decode motor outputs from electrocorticogram (ECoG) signals collected from three monkeys. The dataset was pre-collected and is publicly available (Chao et al., 2010); see **Supplementary Information** for details. The model was run separately for each monkey because the location of the ECoG channels varied across monkeys. Here, the variable $X$ is the brain data (power time series across several frequency bands in 32 electrodes, reduced to 10 principal components), and $Y$ is seven degrees of freedom defining the monkeys' arm movement. To maximally focus on the relation between $X$ and $Y$, $\mu^k$ was set to zero, and the covariance matrix was set to be diagonal and non-state-specific (i.e. global). That means that only the regression coefficients are state-specific. The model is therefore:

$$Y_{t,j} \sim N(X_t \, \beta^{k,j}, \sigma_j^2), \qquad j = 1, \ldots, 7$$

Where j indexes the movement degrees of freedom, $\beta^{k,j}$ is a 10 by 1 vector of regression coefficients (one per principal component), and $\sigma_j$ is a standard deviation parameter shared across states. Note that we ran the model for all movement parameters at once, so that each state is characterised by a (10 by 7) regression matrix $\beta^k$. Alternatively, the model could have been run for each movement parameter separately, such that the state dynamics are not coupled across the output variables.

For illustration, **Figure 3A** shows the regression coefficients of a model with four states and low temporal regularisation (i.e., low prior probability of staying in the same state). As observed, the regression coefficients vary substantially between states, suggesting the suitability of a non-stationarity model. **Figure 3B** shows the corresponding FO. To further characterise the model's behaviour, we ran the model 10 times, across different levels of model complexity in terms of both the number of states and the amount of temporal regularisation. For model comparison, **Figure 3C** shows three different metrics: explained variance (averaged across the seven movement variables; non-cross-validated), free energy, and the entropy of the states' FO (see above).

As expected, as the models become more complex (higher number of states, and lower temporal regularisation), they have higher state entropy and can explain more variance. According to the free energy, the optimal value for the number of states is between 5 and 10 for two of the monkeys, and between 10 and 15 for the other monkey (likely because the latter monkey has more data points).

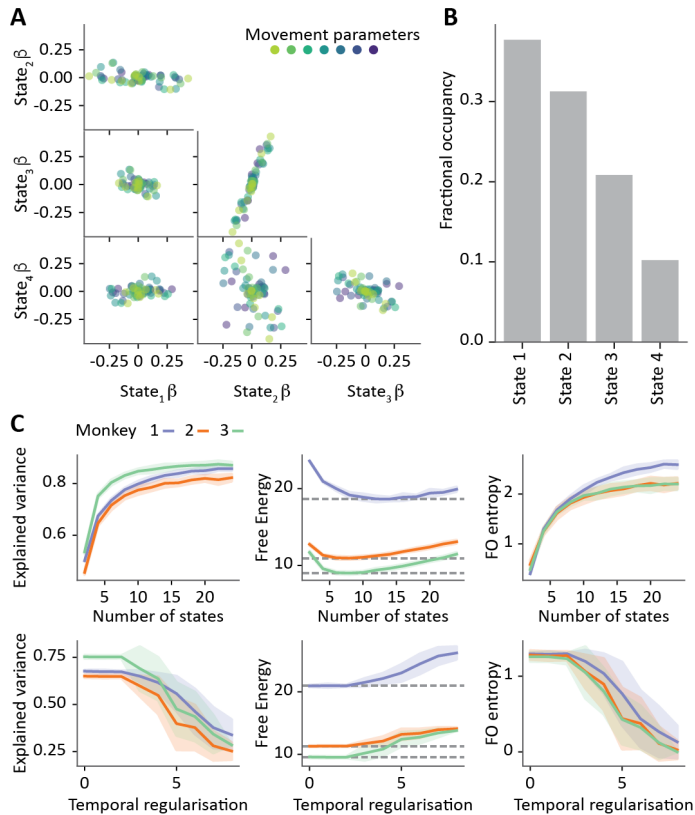

**Figure 3.** Behaviour of the GLHMM on ECoG data collected from three monkeys while they performed free-arm movements. The model was configured such that each state is a regression model of seven movement parameters on the ECoG data. **A)** Scatter plots showing the regression parameters for an example monkey and run. Each colour reflects a different movement parameter, comparing one state against another. **B)** Fractional occupancy (FO), defined as the proportion of time spent in each state. **C)** Given 10 independent runs of the model for each monkey, evaluation of the estimates according to explained variance (averaged across movement parameters), free energy, and FO entropy (a measure of temporal complexity); the models are run across a number of states from 2 to 24 (top panels) and across different amounts of temporal regularisation (bottom panels). For each of the three monkeys, shaded areas represent the standard deviation of the mean, depicted by a thick line. The dashed lines in the middle plots of panel **C)** represent the minimum value of free energy for each monkey.

**Constraining the state paths**

Sometimes, there may be a hypothesis about the structure of the state transitions, or we wish to constrain the solution for practical reasons. For example, in a simple passive visual experiment where images are sequentially shown to a subject, the visual processing cascade is known to follow a series of sequential steps, from low-order processing to higher-order integration (Bullier, 2001). In such cases, it may be useful to force a certain structure in the state sequencing. In the GLHMM, this is implemented by restricting which transitions between states are allowed. For instance, we could enforce that, for each segment of data, the chain of states strictly follows a sequence $k = 1,2,\ldots,K$, while the timing of the transitions is still fully data-driven and trial-specific; or we could have a circular structure, such that it is allowed that state $K$ is followed by state 1; or we could impose that states can only progress forward but not backward; or we could have two blocks of states such that transitions are allowed within one block but not between blocks except for a couple of gating states; etc. See **Figure 4** for some graphical examples.

To illustrate the effects of imposing a structure in the state transitions, we used MEG data from ten participants while they performed a template-matching visual task (Myers et al., 2015). The data is described in detail in the Supplementary Material. Briefly, data was divided in blocks. In each block, participants were presented with one orientation template, which they had to retain in memory. They then viewed a stream of oriented gratings and responded when the presented angle matched the template angle; each stimulus presentation is considered a trial. In this case, we expect that constraining the state paths to be sequential will be a useful description of the data, since the visual processing cascade is likely to follow a stereotyped, feedforward sequence of neural events. The state distribution was set up as a decoding model (Vidaurre et al., 2019), where $X$ is sensor space data reduced from the number of sensors down to 48 principal components, and $Y$ is the distance between the presented and the remembered angle coded (angles were encoded into two covariates using the sine and cosine functions). We imposed a sequential constraint ($k = 1,2,...,K$) and compared the estimates against an unconstrained model in terms of how the state could predict a separate variable: experimental reaction time (under the assumption that a faster transition of states through the sequence should manifest in faster reaction times).

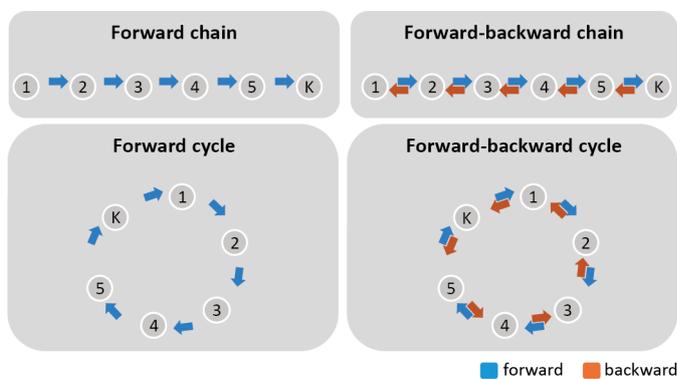

**Figure 4.** Four examples of state transition structures, where only a subset of all possible transitions between states is allowed. Self-transitions are not shown.

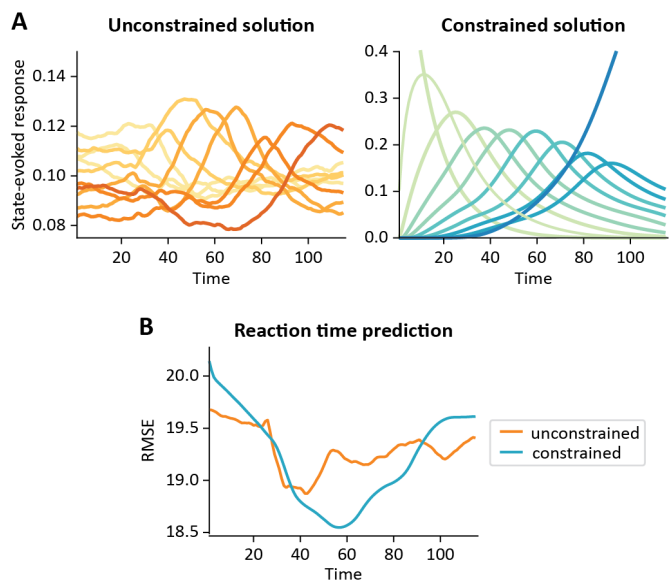

**Figure 5**. Comparing the behaviour of the GLHMM on MEG data during a visual-memory task, between an unconstrained and a constrained (sequential state transition) model. **A)** State-evoked responses for the two models, each with K=10 states. The unconstrained model is represented with warm colours, whereas the constrained model is represented with cool colours; the different states are represented by different shades. **B)** Cross validated, root mean square error (RMSE) of the predicted reaction time, where the state probabilities at a time point were used as predictors of the subject's reaction time for the corresponding trial. Note that a separate predictor was trained for each time point within trial.

**Figure 5** illustrates the estimates in terms of the state-evoked responses (calculated as the average state probabilities across trials; **Figure 5A**), and the root-mean-squared error of predicting reaction time per time point (**Figure 5B**). We observe that, although both the constrained and the unconstrained solution were able to find structure in the data (specifically, by identifying a sequence of processing stages), the unconstrained solution's state-evoked responses were relatively flat in comparison (note the scale in the Y-axes), suggesting that the solution did not effectively describe the various steps of visual processing. Furthermore, the sequential, constrained solution was better able to predict reaction time except for the first and last time points of the trial; this is because, in these time points, the model did not have much freedom to adapt to block-specific variability given the constraint (for example, the first time point is always assigned to state 1, so there is no variability available for reaction time prediction). Overall, this example shows the utility of imposing structure in the transition probability matrix in scenarios where we can expect the process under study to follow a similar structure.

## Assessing brain-behaviour relationships

Here, assessing brain-behaviour relationships refers to exploring connections between an estimated HMM and an external variable not initially considered in the model. Examples include predicting individual traits like age or a cognitive trait from subject-specific HMMs, interrogating the relation between HMM state time courses inferred from neural data and a physiological variable (e.g. pupil size), or finding correlations between brain dynamics and one or more behavioural or cognitive measures across experimental trials. There are two different approaches to interrogating these relationships: inference or statistical testing, and prediction. Statistical testing provides a p-value, while prediction yields (cross-validated) explained variance. Each of these approaches will be described in the following two subsections.

Much of the functionality for statistical testing and prediction share a common structure of variables. Let us assume that $N$ represents the number of observations that we will test across (for example subjects), $p$ is the number of predictors and $q$ is the number of outcomes to test or predict. The test takes a ($N$-by-$p$) design matrix $D$ of independent variables, and a ($N$-by-$q$) matrix $R$ of dependent variables. In situations where there is only one outcome variable ($q$ = 1), $R$ is a vector. Here, $D$ can be a set of HMM-related measurements, and $R$ can be a set of behavioural variables, or vice-versa.

### Statistical testing

Following the estimation of the HMM parameters, the focus here is on finding associations between the HMM estimate and an external variable of interest that was not included in the model. In this context, we can use permutation testing to assess these various types of brain-behavioural associations (Holmes et al., 1996; Nichols & Holmes, 2002), whose main advantage lies in the fact that it does not rely on assumptions about the underlying distribution of the data.

Outcomes, or estimates we wish to assess using permutation testing, include univariate (correlation coefficients) or multivariate (explained variance from regression) statistics. These statistics can be computed from various types of behavioural data, which can be both continuous and categorical. The null hypothesis we are testing is that there is no relationship between $D$ and $R$. In the case of regression, we have a vector of regressors $\alpha$, such that $D\alpha$ is a predictor of $R$. This will yield $q$ p-values (i.e. there are $q$ null-hypotheses). In the case of correlation, we will base the test on univariate correlation coefficients between each column of $D$ and each column of $R$, returning a matrix of $p$-by-$q$ p-values (i.e. there are $p$-by-$q$ null hypotheses). In terms of correction for multiple comparisons, the toolbox offers several different options, including Bonferroni and Benjamini-Hochberg correction.

Finally, deconfounding involves controlling the influence of a third set of variables, $C$, referred to as confounding variables. These confounding variables could potentially act as common drivers affecting both the dependent and independent variables. This process can be addressed by regressing out the confounding variables from both $D$ and $R$ (Pervaiz et al., 2020). This information is summarised in **Table 1**.

**Table 1.** Configuration and options for statistical testing.

| Data Types | Continuous and categorical behavioral data. | |
|---|---|---|
| **Parameters** | $N$: Number of observations (e.g., subjects). <br> $p$: Number of predictors. <br> $q$: Number of outcomes to test. | |
| **Test Design** | ($N$-by-$p$) matrix $D$ of independent variables. <br> ($N$-by-$q$) matrix $R$ of dependent variables. <br><br> $D$ can be HMM-related measurements, and $R$ behavioral variables, or vice-versa. | |
| **Null Hypothesis** | No relationship between $D$ and $R$. | |
| **Outcomes Tested** | Univariate statistics | Multivariate statistics |
| | ($p$-by-$q$) matrix of p-values from the correlation coefficients between $D$ and $R$. | $q$ p-values from regressors ($\alpha$) of $D$ predicting $R$ |
| **Multiple Comparisons Correction** | Returns a matrix of $p$-by-$q$ p-values (i.e., $p$-by-$q$ null hypotheses). | |
| **Deconfounding** | ($N$-by-c) matrix $C$ of confounds, that are regressed out of $D$ and $R$. | |

The HMM-related measurements can either be: (i) the state time courses, representing the probability of each state being active at each time point; (ii) the Viterbi path, providing a categorical assignment instead of a probability; or (iii) a time-aggregated statistic. In the case of the state time courses, a single test is conducted per time point, resulting in either $p$-by-$q$ or $q$ p-values, depending on whether it is a univariate or a multivariate test. This approach allows us to quantify how the brain-behaviour association changes over time. For example, in an experimental paradigm with two conditions presented over multiple repetitions or trials, and with $T$=100 time points per trial, a multivariate test using $D$ as the state time courses (or Viterbi path) and $R$ as the condition type (a single column) would yield $T$ p-values. Alternatively, using a time-aggregated statistic computed across time points provides a single set of ($p$-by-$q$ or $q$) p-values for the entire time span. Some relevant time-aggregated statistics are:

- *Fractional occupancy (FO)*, which, as mentioned, is the total time spent in each state within a certain period of time (e.g., a scanning session).
- *Dwell times*, defined as the average visit duration of a particular state, and thus reflecting information about the temporal dynamics of state transitions.
- *Switching rates*, reflecting the number of transitions between states within a given period.
- *FO entropy*, a measure of uncertainty in the state proportions, with higher values representing a more balanced share of state visits, and zero representing only one state being visited (see the mathematical definition above).

Depending on the experimental design and hypotheses, different permutation schemes can be employed: across-subjects, across-trials, across-sessions-within-subject, and across-visits. These are illustrated in **Figure 6**.

The *across-subjects test* is used to assess the relation between one or more HMM-related aggregated statistics and individual traits. This implies that each observation represents an individual subject, so we shuffle or rearrange across subjects, as depicted in **Figure 6A**. By default, the across-subject test assumes exchangeability across all subjects, indicating that all permutations are possible —i.e., any pair of subjects can be swapped. However, sometimes there might be subjects with familial or other meaningful connections. If ignored, this would violate the underlying assumption of permutation testing, which is that subjects are independent. In such cases, the toolbox can create a hierarchical permutation tree that considers familial relationships between different subjects (Winkler et al., 2015).

The *across-trials test* is used to assess effect differences between trials in one or more experimental sessions. Some examples could be: differences in reaction time between trials in a given task (e.g., does the speed at which the brain traverses a sequence of states lead to faster reaction times?), or differences between two or more experimental conditions (e.g., are there differences in the states' occupancies between a visual and an auditory condition?). As illustrated in **Figure 6B**, we randomly shuffle the order of trials within each session to create a surrogate distribution of the statistic of interest. The test can be applied to the state time courses (having one test per time point in the trial), or to aggregated statistics.

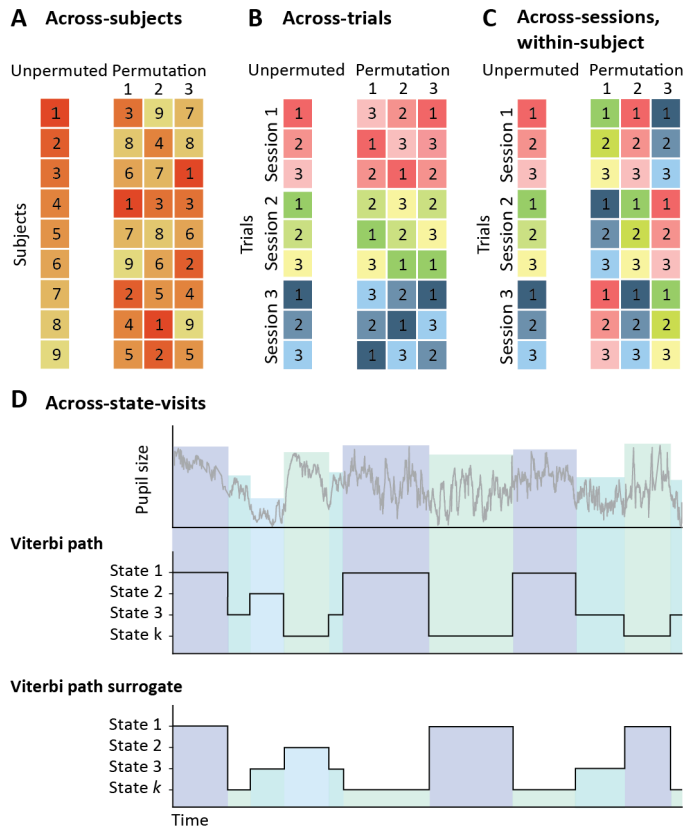

**Figure 6.** Illustration of permutation testing schemes for examining brain-behaviour associations. **A)** Across-subjects: each number corresponds to a subject. **B)** Across-trials: each number corresponds to a trial within a session, and permutations are performed within every session. **C)** Across-sessions-within-subject: each number corresponds to a trial within a session and permutations are performed between sessions, with each session containing the same number of trials. **(D)** Across-visits: based on Viterbi path, a single test is performed for the entire time series (see main text); the test is non-parametrically based on surrogated (random) Viterbi paths, which are randomised paths with similar statistical properties than the original Viterbi path.

The *across-sessions-within-subject* test is used for interrogating variability between different sessions in studies spanning multiple scanning sessions. This is relevant in studies examining changes in brain responses over a longer time period. As illustrated in **Figure 6C**, the across-session-within-subject test involves reshuffling across entire sessions while maintaining the trial order to create a surrogate distribution. Again, the test can be applied per time point on the state time courses or on aggregated statistics.

Finally, the *across-visits* test is used to find associations between the HMM state time courses and a continuously measured variable $S$, such as pupil size or cardiac rate, with as many time points as $D$ or $R$. In the most basic form, there is a single set of ($p$-by-$q$ or $q$) p-values for the entire series. Alternatively, we can also test one state versus the rest (e.g. do activations of state 1 lead to significant changes in $S$?), or one state versus another (e.g. is the value of $S$ significantly different between the activations of state 1 and state 2?). Surrogate Viterbi paths are randomly generated such that the identity of the states is randomised for each visit. This means that, while the surrogate Viterbi paths maintain statistical properties related to the timing of the transition, it introduces randomness in terms of what specific states are visited; see the example in **Figure 6D**.

**Out-of-sample prediction**

Often it is desirable to produce out-of-sample predictions of subject behavioural responses or individual traits from brain data. For example, we might seek to use individual brain activity patterns to classify previously unseen subjects as patients or controls for clinical diagnosis and prognosis, or to predict a cognitive trait from dynamic functional connectivity. The HMM can be used for these types of machine learning problems, enabling us to leverage information from an individual's brain dynamics, beyond mere structural or time-averaged features. In what follows, we will discuss subject-level predictions, but the same principles can be applied to session- or trial-level predictions.

The toolbox includes functions for regression and classification, and routines for both nested cross-validated (CV) prediction (where a single data set is partitioned into folds, using one fold at a time for testing) and prediction given separate train and test data sets (Varoquaux et al., 2017). The CV approach includes an option to account for a group or family structure when creating the CV folds. In this case, subjects will be assigned to CV folds in a way that ensures that related subjects are never split across folds. This is important in some datasets, where there are known relationships between subjects. For example, in the case of the Human Connectome Project (HCP; Van Essen et al., 2013), where many of the subjects are genetically related to other subjects in the dataset, it must be ensured that we are not training the model on one subject and testing on their twin given the well-known strong heritability of certain traits. For prediction, the traits can also be deconfounded by specifying one or more confounds; the deconfounding process can be cross-validated.

There are two different techniques for prediction: the Fisher kernel and a feature-engineering based approach using summary metrics; see **Figure 7A**.

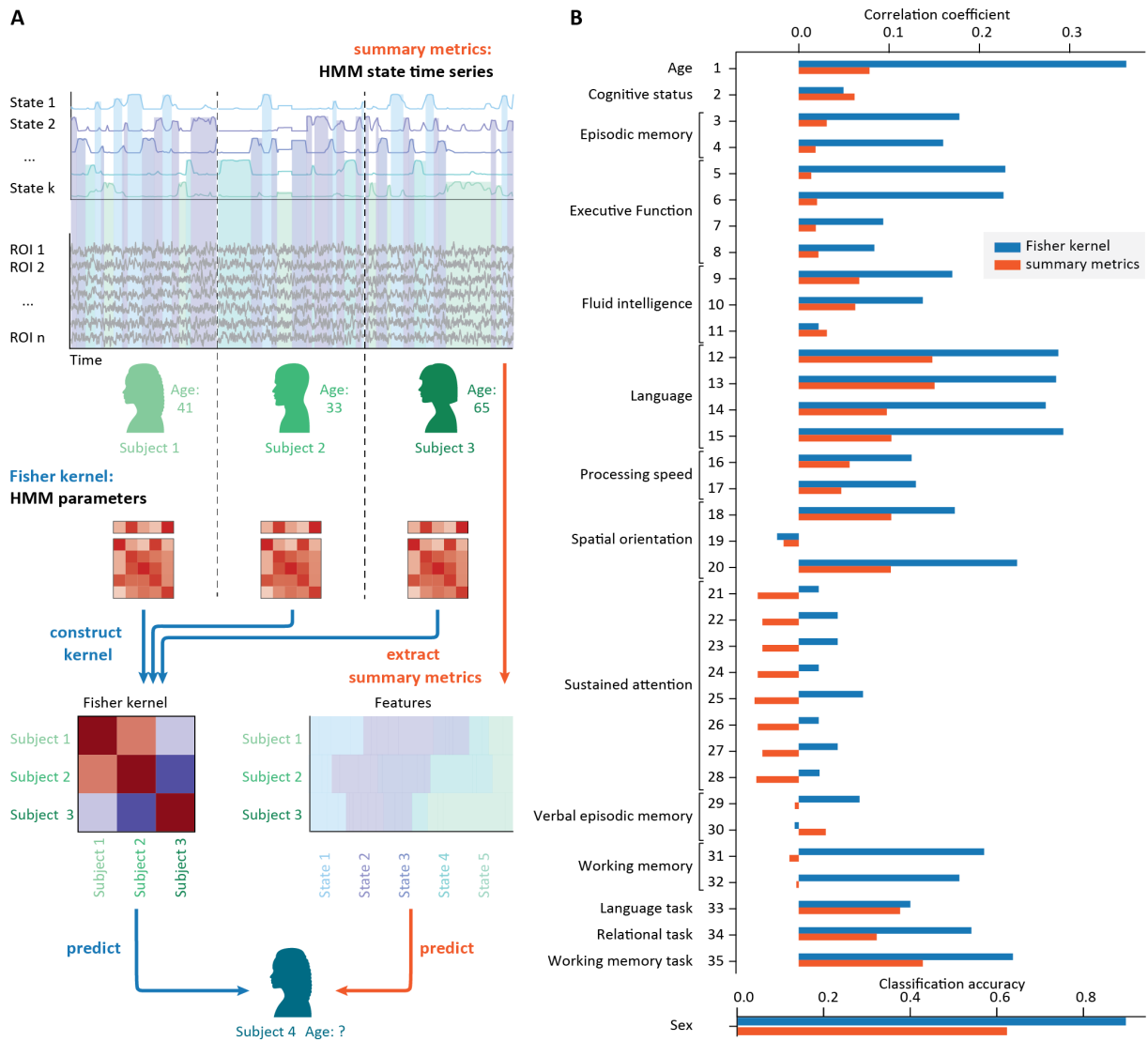

**Figure 7.** Predicting and classifying from an HMM. **A)** Illustration of the Fisher kernel (blue) and the summary metrics (orange) approach to predicting or classifying from an HMM. **B)** Example application in the HCP dataset: Prediction and classification accuracies across predicted variables of both approaches

The Fisher kernel is a mathematically principled and computationally efficient way of predicting or classifying subject variables from HMMs (Jaakkola et al., 1999; Jaakkola & Haussler, 1998), and can result in more accurate and reliable predictions than other methods (Ahrends et al., 2024). Briefly, an HMM is first estimated at the group level (i.e., on all subjects), then subject-specific HMMs are obtained through dual estimation (see section **Dual estimation** above) and then a set of Fisher scores are computed for each subject for the entire set or a subset of parameters of the HMM. For a given subject, the Fisher scores represent how much we would have to change the group-level model to best explain the subject's time series.

Given the resulting score matrix, $D$, and a choice of kernel function, we then compute a similarity metric between all pairs of subjects, building a ($N$-by-$N$, where $N$ is the number of subjects) kernel matrix $K$, which serves as the input to the prediction and classification functions (**Figure 7A,** blue arrows). This kernel function can be configurable as linear or nonlinear. Intuitively, the Fisher kernel function will return a high value for a pair of subjects for which the group-level parameters have to be tuned in a similar way in order to provide good subject-specific models, and a low value for a pair of subjects for which the group-level parameters have to be tuned in a different way. This kernel can be used straightforwardly in any kernel-based prediction model or classifier.

The second option, the feature-engineering approach using summary metrics, extracts pre-defined features from the HMM that can be used for a prediction. The features in this case may be the same as the ones used for statistical testing described above: FO, dwell times, switching rates and FO entropy (**Figure 7A,** orange arrows). Although this method may perform less accurately than the Fisher kernel, it can be advantageous when we wish to explicitly interrogate the predictive power of specific aspects of the model.

As an example, we illustrate the prediction functionality on data from the HCP, where we predict individual traits like age and several cognitive variables, as well as classify sex, from an HMM trained on the resting-state fMRI data. To predict the continuous variables, we used kernel ridge regression for the Fisher kernel and ridge regression for the summary metrics approach (Vidaurre et al., 2021), deconfounding for sex and head motion. To classify sex, we used a support vector machine for the Fisher kernel and logistic regression for the summary metrics. We used 10-fold nested cross-validation, accounting for the family structure. **Figure 7B** shows that, in this dataset, the Fisher kernel approach (blue) generally predicts and classifies at a higher accuracy than the summary metrics approach (orange).

## Affordable training: variational and stochastic variational inference

Given a GLHMM specification, estimation of the parameters is carried out from data using a process called Bayesian variational inference, which aims at minimising a metric called free energy, as explained above. The minimisation is performed by an iterative approximation of the model parameters: (i) the state distribution ($\mu^k, \beta^k, \Sigma^k$), (ii) the initial and transition probabilities ($\pi, \theta$), and (iii) the latent variables ($\gamma, \xi$).

Variational inference is more computationally efficient than methods based on sampling, but it needs to load all the data into memory at once and perform an estimation of all the latent variables in each iteration. For large data sets with upwards of thousands of sessions, like the UK Biobank or the HCP, this can require more memory than is available even on high performance computers. A solution is the use of a stochastic form of variational inference. Stochastic variational inference is a modification where, at each iteration, the parameters are estimated using only a reduced batch of data that is randomly selected. This greatly reduces the memory requirements, because only the selected batch needs to be kept in memory. It also reduces the computation time, since $\gamma$ and $\xi$ (the most time-consuming part of the inference) do not need to be estimated for the whole set of data at each update, but just for the selected batch of data. This means that each update becomes somewhat "noisier", which, counter-intuitively, may sometimes be a beneficial trait compared to standard variational inference (in the sense of producing a solution with lower free energy), as it helps the updates elude local minima thanks to the increased stochasticity.

We tested training performance on resting-state fMRI data from the HCP. **Figure 8** demonstrates the performance of the standard and the stochastic inference for different batch sizes (5, 10, 20, 100, 200, 500, 800 and for comparison, 1003, which is the size of the entire dataset) for 10 runs. **Figure 8A** shows the relationship between the free energy and number of training cycles until convergence for standard (non-stochastic) variational inference and for stochastic variational inference across batch sizes.

In terms of free energy, non-stochastic training generally outperformed stochastic training in this case, with performance on average improving as batch size increases. However, the largest differences occur between smaller batch sizes >10%, with larger batch sizes showing diminishing returns. **Figure 8B** shows, for the same runs, the number of Floating Point OPerations (FLOPs; a load-insensitive metric of computing time) per cycle. As observed, smaller batch sizes have notably smaller per-cycle computational demands than larger batch sizes. Although not shown here, stochastic variational inference has also a benefit in terms of memory usage since it does not need to load the entire data set at once.

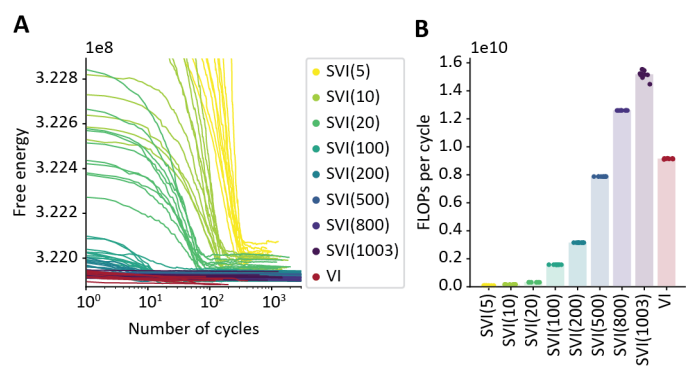

**Figure 8.** Overview of the GLHMM training performance, comparing standard (non-stochastic) variational inference (VI) and stochastic variational inference (SVI) across various batch sizes (see main text). Each type of inference was run 10 times. **A)** Free energy vs. number of cycles until convergence (excluding initialisation routines). **B)** Number of FLOPs per cycle by training settings.

## Discussion

The HMM is a conceptually simple yet powerful framework for characterising time series data, with the benefit of being based on an unambiguously specified generative model. In this paper, we have presented the GLHMM approach and related software. Essentially, the GLHMM is an HMM that uses linear regression to parameterise the Gaussian state observation models. This can be adapted for a wide range of uses that are common in electrophysiology and neuroimaging, including unsupervised and encoding/decoding modelling. We have here illustrated its use on fMRI, MEG, ECoG and LFP data.

We can describe brain data at various levels of complexity, from simple correlations or averaged spectral features, to very complex biophysical models. The GLHMM can be considered as a model of intermediate complexity that allows for effectively testing hypotheses and finding relations to behaviour. Therefore, the assumptions of the HMM (a discrete number of states that are Markovian) should not be considered as representing a biological ground-truth, but as an interpretable and mathematically convenient description. For example, a finite latent model like the HMM may be a good approximation of a continuous generative process, even if there is nothing discrete in this process. This is also relevant when we think about model selection: which is the best model from a Bayesian criterion point of view (i.e. according to the free energy) is a practical question, and should not be interpreted biologically (Vidaurre, 2023).

Importantly, the inference of the model has the potential to capture other aspects of the data, even if these are not explicitly modelled. For example, a common misconception is that the HMM cannot infer state timecourses with long-term dependencies. While it is true that the Markovian prior only captures short-term dependencies, it does not actually penalise the fitted state time courses if they have long-term dependencies. The presence of long-term dependencies in HMM-inferred state time courses can be clearly demonstrated empirically (e.g. see the Fano Factor plots of state activation rates in Higgins et al, 2021). Given this, and since the GLHMM does not make brain-specific biophysical assumptions, it can readily be applied to other domains with time series or other sequential data.

Altogether, the introduced Python toolbox stands as a versatile framework that allows modelling time-varying interactions within and between time series with an emphasis on subsequent prediction and statistical testing. This enables users to evaluate the GLHMM against variables not initially included in the model. These variables can be examined at different levels, including the subject level (e.g. the relationship between brain dynamics and age), the trial level (e.g. whether brain dynamics differ between two experimental condition), and the time-point level (e.g. with respect to a continuously recorded variable such as physiological measures like pupillometry). This framework can be applied to both basic and clinical research questions, as well as brain or non-brain data to describe spatial, spectral, and temporal patterns in time series data, and assess their relations to behaviour.

**Code:** The developed toolbox, which can be installed from TestPyPi, is publicly available at: https://github.com/vidaurre/glhmm
The code for this paper is publicly available at: https://github.com/CFIN-analysis/GLHMM_paper

**Acknowledgements**: DV is supported by a Novo Nordisk Foundation Emerging Investigator Fellowship (NNF19OC-0054895) and an ERC Starting Grant (ERC-StG-2019-850404). MWW's research is supported by the NIHR Oxford Health Biomedical Research Center, the Wellcome Trust (106183/Z/14/Z, 215573/Z/19/Z), the New Therapeutics in Alzheimer's Diseases (NTAD) study supported by UK MRC, the Dementia Platform UK (RG94383/RG89702) and the EU-project euSNN (MSCA-ITN H2020–860563). This research was funded in part by the Wellcome Trust (215573/Z/19/Z). For the purpose of Open Access, the author has applied a CC BY public copyright licence to any Author Accepted Manuscript version arising from this submission. We wish to thank Lennard Krause for technical support.

## Supplemental information

**Data specifications**

**ECoG, monkey brain-computer interface data.** This dataset was pre-collected and made publicly available. Data were first presented in (Chao et al., 2010). The ECoG signals were recorded at a sampling rate of 1 kHz per channel, for a total of thirty-two electrodes implanted in the right hemisphere, and band-pass filtered from 0.3 to 500 Hz. We considered a time-frequency representation of the ECoG signals containing 1600 variables (32 electrodes, 10 frequency bins, and 5 time lags), from which we extracted 10 principal components, and downsampled to 250Hz. The monkey's movements were captured at a sampling rate of 120 Hz, and upsampled to 250Hz to match the sampling frequency of the brain data. We used data from three different monkeys, with one continuous scanning session of 2991, 1499 and 1499 time points after sampling, respectively.

**Magnetoencephalography, visual-memory task.** Neuromagnetic data were acquired using a whole-head VectorView system (204 planar gradiometers, 102 magnetometers; Elekta Neuromag). The signals were sampled at a rate of 1000 Hz and online band-pass filtered between 0.03 and 300 Hz. The raw MEG data were visually inspected for artefacts, de-noised and motion-corrected, and downsampled to 250 Hz. Artefacts arising from eye blinks and heartbeats were removed via independent component analysis. Epochs were generated around each stimulus onset (from 0 to 0.6 s) and visually inspected to eliminate any remaining trials with excessive noise. The task consisted of eight brief (approximately 6 min) blocks, in which 480 stimuli were presented (resulting in a total of 3840 stimulus presentations per session). Each block began with the presentation of a target orientation (drawn at random, without replacement, from the 16 stimulus orientations), displayed centrally as a green line. The stimulus stream consisted of randomly oriented Gabor patches, presented centrally for 100 ms, at an average rate of 650 ms. Stimuli had 16 possible angles (5.625–174.375°, in steps of 11.25°). Participants were instructed to respond whenever a Gabor patch with a matching orientation appeared. Since stimuli were drawn uniformly from the 16 possible orientations, 1/16 of all stimuli were targets. The angles were encoded into two covariates using the sine and cosine functions, plus some Gaussian noise for model inference stability. Each block was cut into three shorter segments, giving participants brief rest periods. During the rest periods, the target orientation was presented again as a reminder. Participants were instructed to respond as quickly and accurately as possible.

**Human Connectome Project: resting-state fMRI and behavioural data.** The resting-state fMRI and behavioural data are available from the Human Connectome Project (HCP) database at db.humanconnectomeproject.org and described in detail in (Van Essen et al., 2013) and (Smith et al., 2013). Briefly, we here used resting-state fMRI and behavioural data from 1,001 subjects from the HCP S1200 release. The fMRI data were collected in a 3T MRI scanner at 2 mm isotropic spatial resolution and a repetition time (TR) of 0.72 seconds over four separate scanning sessions of 14 min. 33 sec. each, which we here concatenated, resulting in 4,800 time points per subject. We used time series in the groupICA50 parcellation. For the predictions, we deconfounded for sex and head motion.